# An Integrated Framework to Recommend Personalized Retention Actions to Control B2C E-Commerce Customer Churn

Shini Renjith

*Assistant Professor, Department of Computer Science & Engineering*
*Sree Buddha College of Engineering*
*Pattoor, Alappuzha, Kerala, India*
shinirenjith@gmail.com

*Abstract - Considering the level of competition prevailing in Business-to-Consumer (B2C) E-Commerce domain and the huge investments required to attract new customers, firms are now giving more focus to reduce their customer churn rate. Churn rate is the ratio of customers who part away with the firm in a specific time period. One of the best mechanism to retain current customers is to identify any potential churn and respond fast to prevent it. Detecting early signs of a potential churn, recognizing what the customer is looking for by the movement and automating personalized win back campaigns are essential to sustain business in this era of competition. E-Commerce firms normally possess large volume of data pertaining to their existing customers like transaction history, search history, periodicity of purchases, etc. Data mining techniques can be applied to analyse customer behaviour and to predict the potential customer attrition so that special marketing strategies can be adopted to retain them. This paper proposes an integrated model that can predict customer churn and also recommend personalized win back actions.*

## I. INTRODUCTION

B2C E-commerce is witnessing a step by step growth in terms of its contribution to the worldwide retail market. According to Forrester Research, as on 2014 E-Commerce is contributing 5.9% of the overall retail revenue amounting to $1.3 trillion. E-Commerce share is expected to grow to $2.5 trillion by 2018 reaching 8.8% contribution to retail revenue. While E-Commerce is having such a great growth potential, there are many challenges for the firms in this area. In a seller perspective customer retention is the key for their profitability. Most of the initial transactions are not profitable in B2C E-Commerce business while accounting the average cost for new customer acquisition. A customer relationship will start making profit only after a few repeated business, when the average cost per serving the customer comes down. This underlines the need of customer retention and reduction of customer churn.

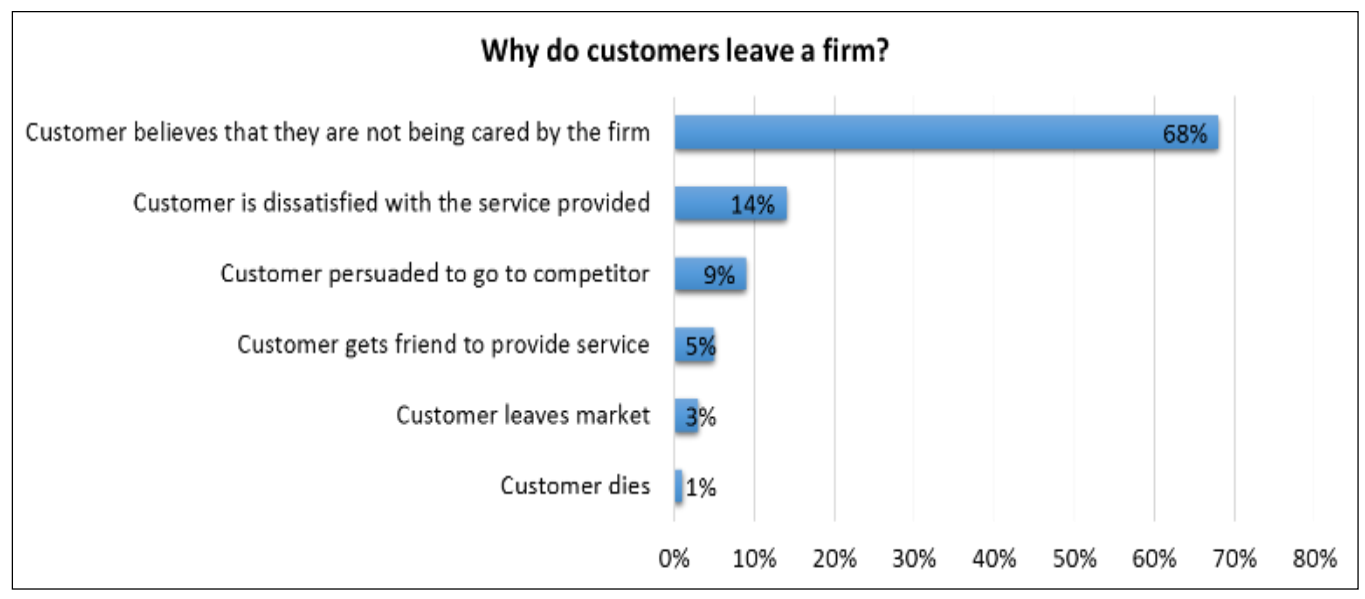

Figure 1. Reasons for Customer Churn

Every E-Commerce business house understands that only a part of their customers stick with them, while the rest stop shopping after one or two transactions. According to a survey conducted by the Rockefeller Foundation, majority of customers mentioned that they move on from sellers as they are not being cared [Fig. 1]. The real challenge is to identify which customer is moving to the competitor and which customer is going idle for a while. Historical data can be mined in such a scenario to identify different purchasing patterns of customers. For example, some shoppers might do their purchases on every alternate weeks, some might do near to their salary day and some might purchase only around holidays. This has a dependency on what is the item being purchased as well. But when a consumer deviate from the purchase pattern which was observed for a while, it may be a potential case of churn.

Once a potential parting away customer is identified, the retailer can try out personalized retention strategies. Individualized offers may do excellent results in such scenarios. For example, a customer who orders a particular brand products on every school opening, but shows signs of parting away with recent purchase patterns may get excited and come back when he get a call/intimation from the seller on the availability of his preferred items at a discount rate.

The proposed platform attempts to resolve both the aspects of customer retention – early detection of customer churn and timely personalized actions to stop the churn. Subsequent sections of this paper deals with the details of existing works in the related areas and the discussion on the proposed solution framework.

## II. RELATED WORKS

There are many researches in the past trying to tackle the challenge of accurately predicting potential customer churn with the help of statistical, data mining and/or machine learning strategies. In these studies, once any candidate churn case is identified it is left for the managerial decision to take appropriate retention action, which is normally constrained by their knowledge level. However these actions are very critical in terms of achieving the end objective – the customer retention. Adding a personalized aspect to the retention action by analyzing the customer preferences is a rarely attempted area.

Scholars have considered various aspects in predicting potential customer churns. Hennig-Thurau et al. [1] approached the churn issue across industries with a customer satisfaction angle whereas Bolton et al. [2] focused on the loyalty program effectiveness. Neslin et al. [3] and Jamal et al. [4] have looked at marketing models to detect probable parting away customers. Hadden et al. [5] attempted the churn prediction with the help of data mining models. Specific prediction models could be found for customer churn in industries like financial services [6] and telecommunication [7], [8], [9]. Adomavicius et al. [10] reviewed different personalization aspects of recommendation systems and explained an integrated personalization process in their study.

Most of these studies are in the area of web content personalization and recommender systems creation. The key parameter for consideration in these cases are the individual customer profile which represents the personal preferences. In a recommender system context, the other aspects that comes to consideration are set of ratings (trust factor) [11] and/or customer's past transaction traits [12], [13], [14]. The recommendation is generated by analyzing the similarities [15] between customer profiles who bought or rated a product and those of customers who did not. In the case of web content personalization, the web usage statistics form the user profile and the recommendation is generated by associating a content with similarities with the user's profile [16].

## III. METHODOLOGY

This paper proposes a simple platform to manage customer churn in terms of prediction and recommendation on win back strategies [Fig. 2]. This framework consists of three stages in the approach – churn detection, customer profiling and recommendations for timely personalized retention actions.

The most effective technique to control customer attrition is to understand the customer and utilize this understanding to ensure customized customer service and maintain a positive relationship. This can be achieved only by knowing or examining the customer profile and behavior. Analytics helps in this context to act proactively by providing a view to customer behavior by mining and/or analyzing the historical data. Analytics can help businesses to convert their objectives into data driven problems, so that each of them can be addressed specifically to arrive at the final set of action items/recommendations. Analytical techniques like predictive modeling and customer profiling can be leveraged as powerful tools to manage the problem of customer churn.

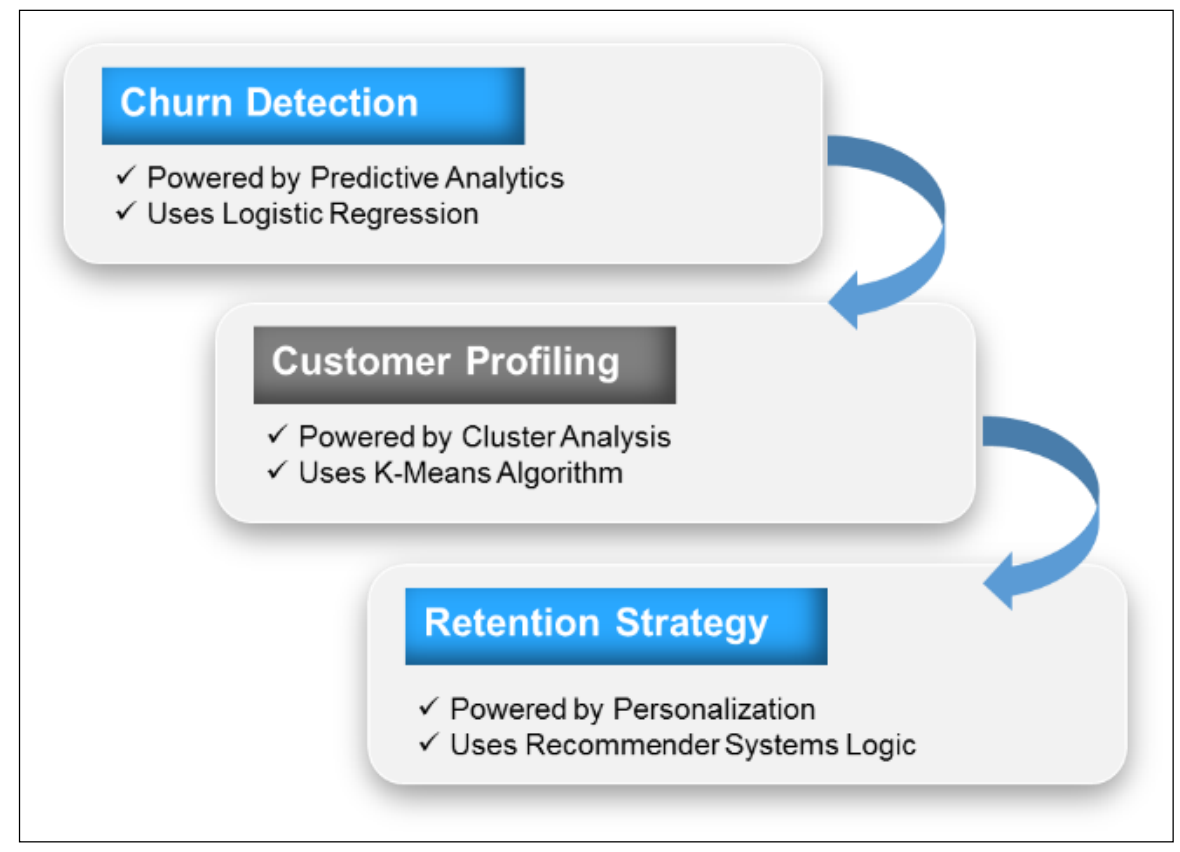

Figure 2. Proposed Customer Retention Platform

### A. *Churn Detection via Predictive Analytics*

Predictive analytics leverages statistical models or machine learning techniques to detect probable attrition behavior of a customer by populating risk scores. In this stage, the customer profile and transactional behavior [Fig. 3] are examined and used for predicting the probability for attrition. The demographic and transactional data collated for each customer form the input for the prediction process. An additional input that can be taken into consideration is the Net Promoter Score (NPS) which is a measure of customer loyalty for the company.

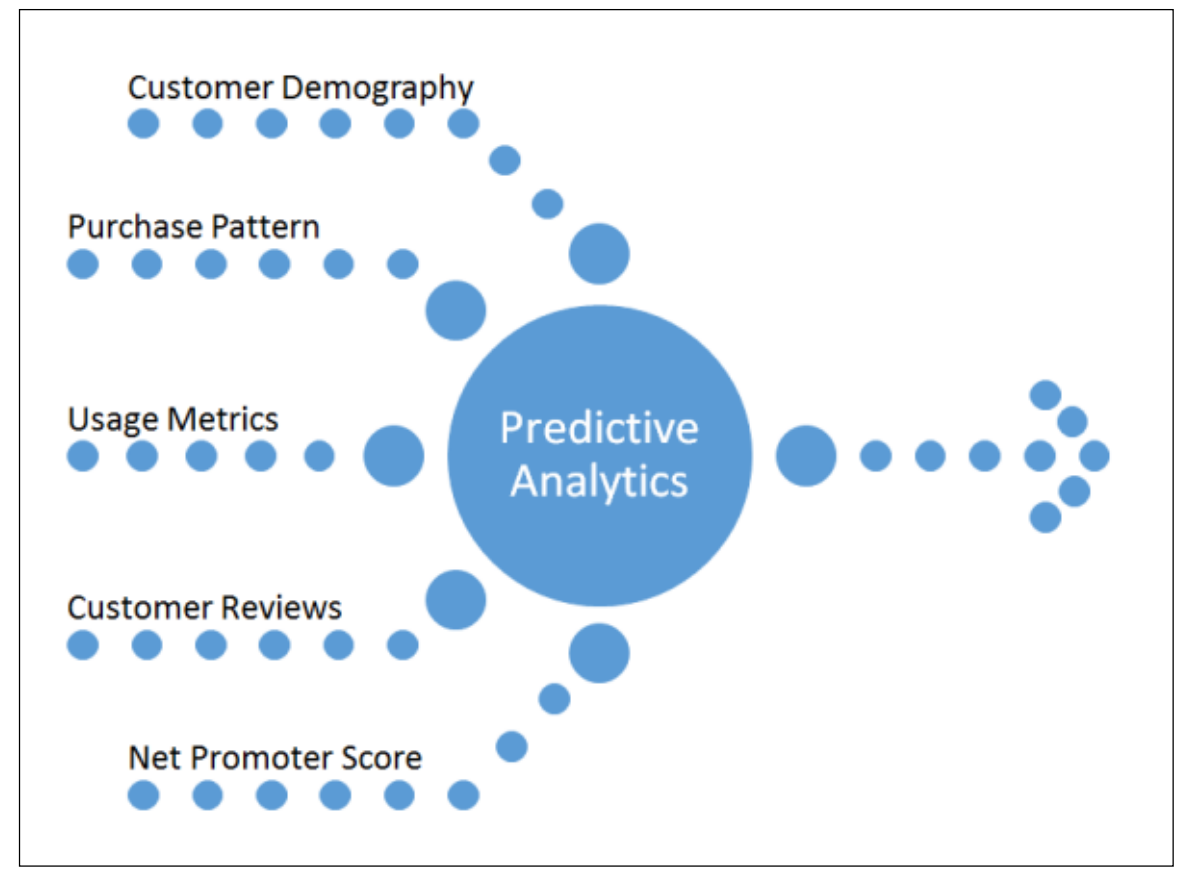

Figure 3. Inputs for Predictive Analyis

Either regression techniques or machine learning techniques can be used to conduct the predictive analytics. Regression techniques focus on firming up a mathematical model to represent the interactions between different input variable in consideration. There are different regression models that can be leveraged depending on the context where predictive analysis need to done. Logistic regression can be used in B2C customer attrition context to build the predictive model [17]. The predictive scores generated through this model will represent the risk quotient for each customer towards attrition. Logistic regression is explained in the book [18] by J.S. Cramer. The normal regression model with n inputs can be represented by the formula:

$$P(X) = \alpha + b_1x_1+b_2x_2+...+b_nx_n \quad (1)$$

Corresponding formula in the logistic regression model is:

$$Q(X) = \frac{1}{1+e^{-(\alpha+b_1x_1+b_2x_2+...+b_nx_n)}} \quad (2)$$

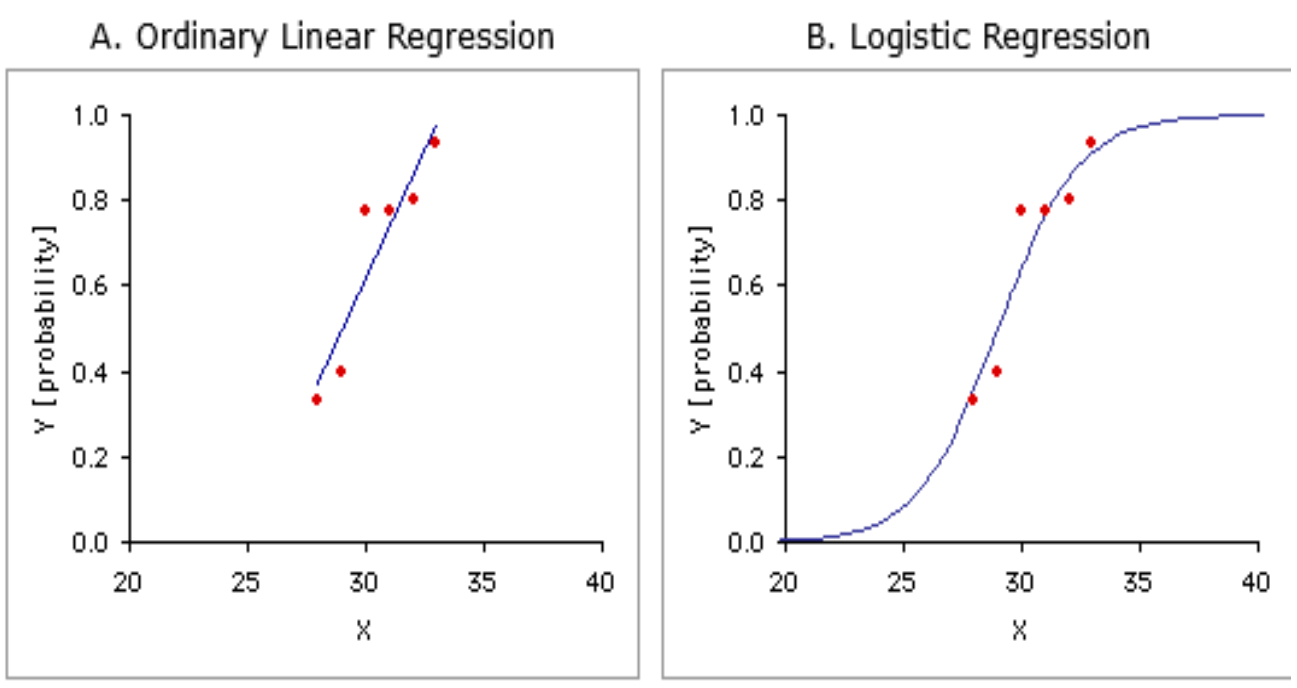

Figure 4. Linear Regression vs Logistic Regression

Fig. 4 shows the difference between linear regression and logistic regression plotted using same data set with observed probabilities Y, on the independent variable X. In linear regression, if the regression line is extended a few units upward or downward along the X axis, the predicted probabilities will fall outside legitimate range of 0.0 to 1.0. Logistic regression fits the correlation between X and Y with an S-shaped curve which is mathematically constrained to fit within the range of 0.0 to 1.0 on the Y axis.

In the B2C churn scenario, based on the predictive risk score the customers who are more likely to churn can be segmented out. The risk score along with other customer details available will form the input data for the next stage - cluster analysis to segment customers and apply strategies.

### *B. Customer Profiling via Cluster Analysis*

Customer profiling is performed through cluster analysis based on the motives for attrition. Cluster Analysis is used to partition the complete customer population based on their exhibited transactional behavior and demographic information. Fig. 5 demonstrates the different steps involved in cluster analysis:

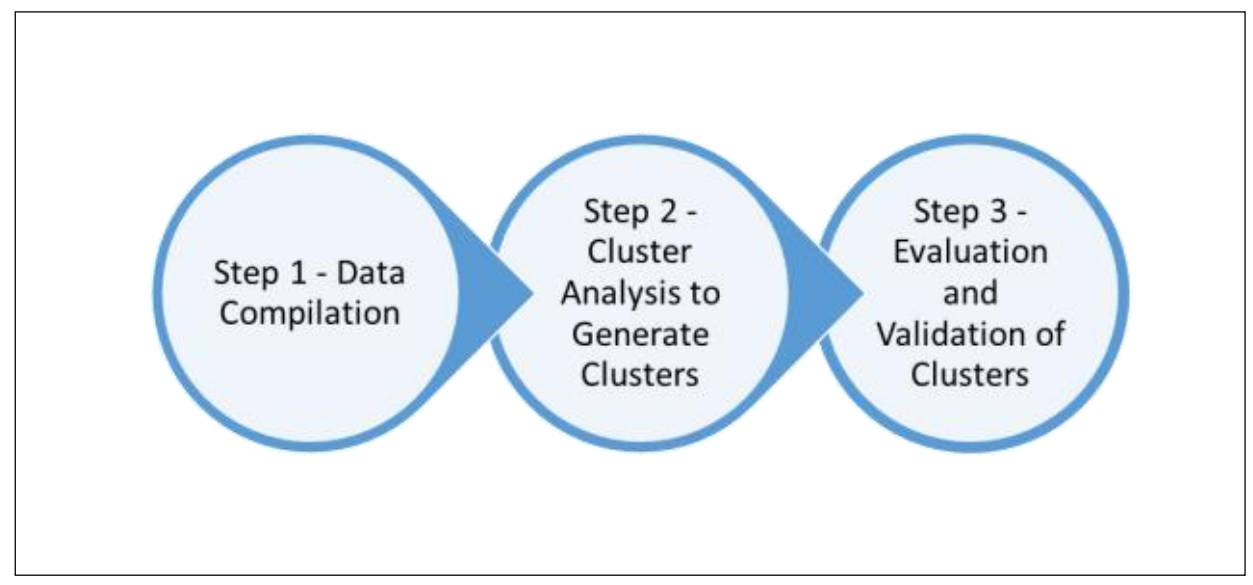

Figure 5. Customer Profiling - Stages

- Data Compilation – Data Compilation activity comprise of collation of input data from different dimensions [Fig. 6]. like behavioral, demographic, transactional, etc. The historic data illustrating the cause of attrition for customers is also accounted for construction of clusters.

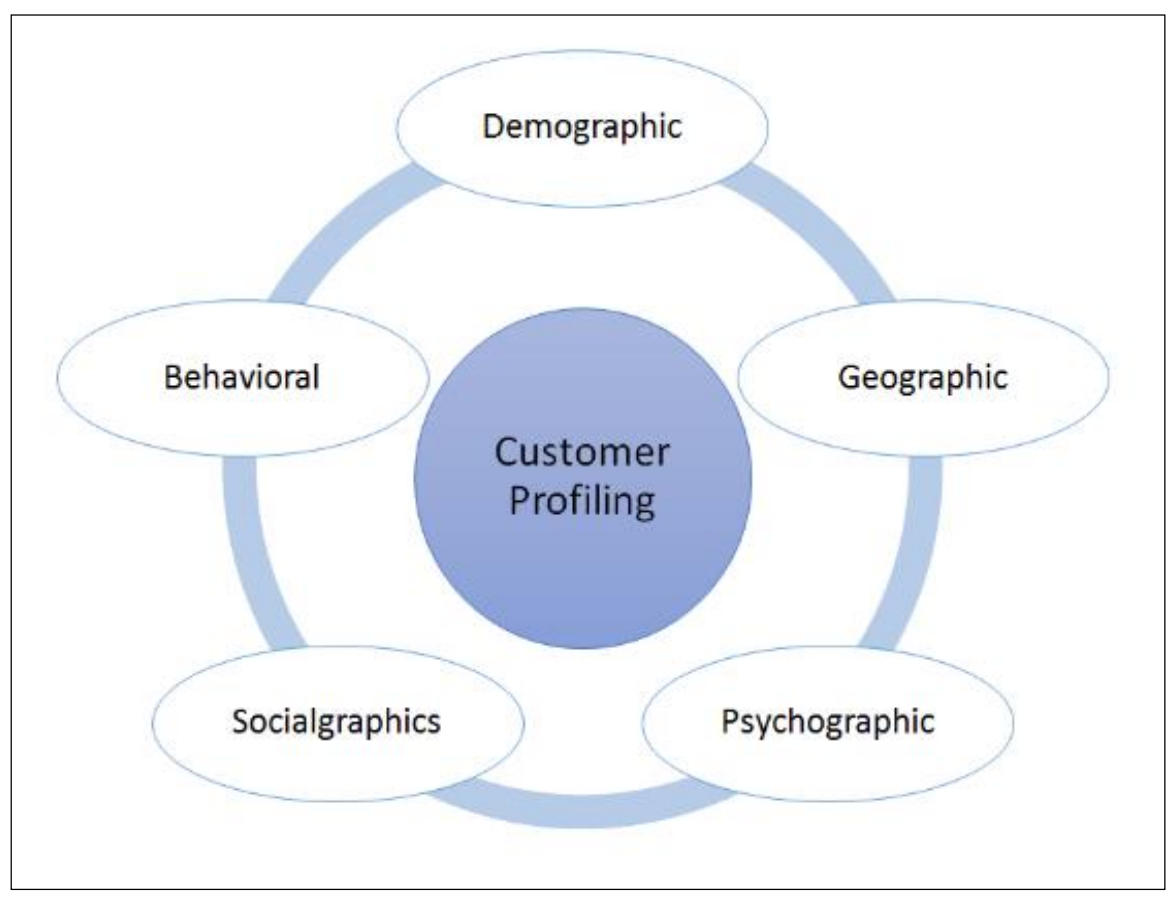

Figure 6. Customer Profiling - Dimensions

- Cluster Analysis to Generate Clusters – Cluster Analysis is performed based on the k-means clustering algorithm [19], [20] which runs on an iterative mode. Through iteration, each data point is assigned to a cluster based on the minimum Euclidean distance from the k-cluster centroids. For a set of n observations ($x_1$, $x_2$, …, $x_n$), k-means clustering aims to split these n observations into k ($\leqslant$ n) disjoint sets S = {$S_1$, $S_2$, …, $S_k$} so as to minimize the within-cluster sum of squares. Mathematically, its objective is to find

$$j = \min \sum_{i=1}^{k} \sum_{x\in S1} \|x-\mu i\|^2 \quad (3)$$

where $\mu\iota$ is the mean of points in $S_i$.

The cluster centroids will get fine-tuned as the data points in each cluster get updated based on minimum distance calculation. The algorithm used for cluster analysis is depicted in [Fig.7]

1. Identify k points in the space where the objects that need to be clustered are represented. These k points correspond to initial group centroids.
2. Assign each object to the group that has the nearest centroid.
3. When all objects are assigned, re-calculate the k centroid positions.
4. Repeat Steps 2 and 3 until the centroids no longer require update.

Figure 7. k-means algorithm

- Evaluation and Validation of Clusters – The generated clusters need to be validated for accuracy and practical application. Evaluation of a clustering solution typically involves:

(1) Verification of each segment size
(2) Evaluation of the cluster centroids and creation of descriptive profiles for each cluster in terms of the segmentation variables, and
(3) Comparison of clusters with regard to variables that are not used in clustering process

### C. Personalized Retention Strategies

The resulting cluster profiles from cluster analysis represent the behavioral and demographic characteristics of customers. Each cluster represent a unique collection of characteristics and these can be leveraged in strategic decision making for preventing customer churn. In simple terms, targeted retention or win back campaigns can be planned on a case to case basis for each clusters.

The low risk customers can be offered a loyalty program [21] – like a special discount offer or redeemable reward points- to ensure retention. The customers who are more prone to attrite need to be handled in a more personalized manner. Various approaches that can be adopted to generate personalized actions include:

- Top-down Approach – Based on managerial level experience from prior actions, actions are defined independent of customer profiles.
- Bottom-up Approach – Customer profiles are defined prior to defining actions, but requires considerable amount of supervision effort.
- Customized Approach - The offers has to be granular enough and the customers are free to choose from the wide set of alternatives.
- Similarity-based Approach - Actions are triggered based on customer preferences inferred from customer profiles

The key criteria for finalizing the approach for retention strategies are (i) better targeting and (ii) control on marketing process. Bottom-up and Customized approaches lacks control in terms of marketing process. Top-down approach fall short in terms of targeting. This lead to the choice of Similarity-based Approach for generating personalized retention actions.

Recommender Systems are the most common example of Similarity-based Approach. It helps in recommending items and/or services to customers based on the perceived utility of it to the customer. These actions are highly personalized as the set of items recommended to each customer is different and is based on his/her preferences and past transaction patterns. Recommender Systems estimate the utility of an item by using a similarity function which may vary depending on the recommendation approach adopted among:

- Content-based [22] - Based on past transactions and/or preferences of the same customer, similar items are recommended
- Collaborative [23] - Based on past transactions and/or preferences of customers with similar preferences and tastes, similar items are recommended
- Hybrid [14] - Combination of Content-based and Collaborative methods.

Following are assumptions / hypothesis considered for proposing this personalized retention strategy based on recommender systems.

- A potential customer attrition can be arrested to a considerable extent by recommending relevant offers / services to them.
- Relevancy of an offer / service may vary with customer, ie. same recommendation cannot control the churn of all customers. Behavior of loyal customers with same taste can be looked at for identifying the relevant offer / service to be offered to a customer who may part away.
- Customers who behave in same manner have similar preferences.

Based on the above assumptions, the proposed model adopt collaborative filtering approach for generating recommendations to carry out personalized retention strategies. Each risky customer will be recommended with offers / services that loyal customers with similar preferences and tastes liked in the past. The hybrid model is not attempted as the same customer may exhibits different purchase pattern based on the different demographic stages he/she is currently undergoing like life cycle stages, geographic locations, etc.

In this model, each customer is represented in a vector space which cover his/her demographic and transactional data. This will help determining the similarity between users in terms of their demographic profiles and behavior patterns. A similarity function can help us in determining how similar two customers are – using Cosine similarity measure, a high value indicates that the customers are very similar. So we identify a neighborhood for each of the risky customers so that recommendations can be proposed. The neighborhood can be defined as a set of customers, j whose preferences and tastes are matching to a customer, i formed by applying cosine similarity measure. Mathematically it can be represented as below

$$Sim(i,j) = \cos(i,j) = \frac{i.j}{\|i\| \times \|j\|} \quad (4)$$

$$\cos(i,j) = \frac{\sum_{s \in S_{ij}} r_{i,s} r_{j,s}}{\sqrt{\sum_{s \in S_{ij}} (r_{i,s})^2} \sqrt{\sum_{s \in S_{ij}} (r_{j,s})^2}} \quad (5)$$

where

$r_{i,s}$ is the rating of an item s by user i
$r_{j,s}$ is the rating of an item s by user j
$S_{ij} = \{s \in Items | r_{i,s} \neq \emptyset \wedge r_{j,s} \neq \emptyset\}$, set of all items rated by both user I and user j
$i.j$ is the dot product of vectors i and j

In this case customer i is part of the predicted set of potential churners and customer j is part of the loyal customer set J. Based on the predictive analysis performed in the previous step, customers with more than 90% probability in terms of loyalty form the set J.

The core logic behind the recommender system concept is that the behavior of similar customers in a given context will be of similar nature. So a potentially risky (prone to churn) customer can be retained by providing offers/services similar to a loyal customer who is having similar preferences and tastes (who is in the neighborhood as per similarity tests). The recommendation algorithm will capture these detail and generate the list of offers / services that should be presented to each customer at risk.

The key difference between a traditional recommender system and the customer retention framework is in the area of neighborhood formation. In a traditional recommender system the neighborhood is formed with all similar customers to the customer in consideration. In this framework, the neighborhood is selected from a subset of all customers, which is the set of all loyal customers (ie, customers with high loyalty).

## IV. DISCUSSION

The framework being proposed is a three tier approach to detect and control customer churn. As the first stage logistic regression is used to identify the predictive risk score for each customer. The dependent variable in this case is whether the customer under analysis is a potential churner or not. There is no restriction in the number of independent variables that can be considered, but it is required to ensure that each input variable is really independent of all others under consideration. Failing to ensure this prerequisite, the framework will tend to give more weightage to those inputs which are related to each other and may mislead in churn prediction.

Once the potential churners are identified, customer profiling is carried out using k-means clustering. Each customer group formed as part of this step qualify for unique retention strategies appropriate to their profiles and behavioral pattern. The two learning strategies considered are k-means and hierarchical clustering algorithms. K-means algorithm is chosen in the model considering its run-time efficiency. The performance of hierarchical clustering algorithm decreases as the number of records under consideration increases. There is an increase in execution time for K-means algorithm also but the performance is better in comparison with hierarchical counterpart.

As the final step, collaborative filtering mechanism is adopted to extract the details of promotions liked by loyal customers who were having similar preferences. Content based filtering model is not considered as it deals only with past actions of the same customer which is having no meaning in churn and retention context. Hybrid model is also excluded from consideration as a user may demonstrate varying purchase patterns during different demographic stages.

## V. CONCLUSION

As per multiple studies among online retailers, attracting a new customer is five times costlier than keeping an existing one. According a study conducted by the management consulting firm Bain & Co., if a firm could achieve 5% increase in the customer retention rate, it can result in a minimum of 25% increase in their profit. Marketing metrics on probability of selling indicates that existing customers may do more business (60-70% of probability) in comparison with a new prospects (5-20% probability). All these statistics point toward the importance of reduction of customer churn.

The framework discussed in this paper helps sensing the potential customer attrition with the help of predictive analytics done on the huge amount of data the retailers hold at any moment. Further it suggests appropriate retention strategies based on the customer profile and traits observed in past transactions. Hybrid recommendation models are leveraged to ensure that the social aspect and latest trends among the similar demographic groups are being accounted for.

As a next step, the framework can be extended further to measure the success rate of the retention strategies. This will help the system to learn better using artificial intelligence and/or machine learning techniques and thereby improve the recommendation process. Another potential expansion area is offering proactive communications and/or recommendations to the customers with new trends observed in the same demographic classes.

## ACKNOWLEDGMENT

I acknowledge and would like to thank Mr. Renjith Ranganadhan for the support, guidance, reviews, valuable suggestions and very useful discussions in the domain of retail marketing.